 \documentclass[prb,preprint]{revtex4}

 \usepackage{dcolumn,amsmath}
 \usepackage{bm}
\def\veps{\varepsilon}
\def\etal{{\it et al.}\ }

\begin{document}

% {\em Ab initio}
% \title{Relativistic correlation 
%        study of spectroscopic constants in E112H and its cation: 
%        Whether E112 is relatively inert element?}
 \title{Is E112 a relatively inert element? Benchmark relativistic
        correlation study of spectroscopic constants in E112H and its cation.}

\author{N.S.\ Mosyagin}\email{mosyagin@pnpi.spb.ru}
                       \homepage{http://www.qchem.pnpi.spb.ru}
\author{T.A.\ Isaev}
\author{A.V.\ Titov}

\affiliation{Petersburg Nuclear Physics Institute, 
             Gatchina, St.-Petersburg district 188300, Russia}

\date{\today}

\begin{abstract}
 We report the 
% "Phys.Rev preferes not to make claims on novelity or priority..."
% see their corrections to PbO-CI article
 first results of relativistic correlation calculation of the spectroscopic
 properties for the ground state of E112H and its cation in which spin-orbit
 interaction is taken into account
% iteratively at the correlation stage.
 non-perturbatively.  Studying the properties of E112 (eka-Hg) is required for
 chemical identification of its long-lived isotope, $^{283}$112.
% The methods of generalized relativistic effective core potential, both
% relativistic and nonrelativistic Fock-space coupled cluster with single and
% double cluster amplitudes and spin-orbit direct configuration interaction
% Most advanced {\em ab initio} methods are used to provide high accuracy.  
 It is shown that appropriate accounting for spin-orbit effects leads to
 dramatic impact on
 the properties of E112H whereas they are not so important
 for E112H$^+$.  The calculated equilibrium distance,
 $R_e^{\rm calc}{=}1.662$\,\AA, in E112H is notably smaller than
 $R_e^{\rm expt}{=}(1.738\pm0.003)$\,\AA\ and $R_e^{\rm calc}{=}1.738$\,\AA\ 
% (studied within similar approximations) 
 in HgH, whereas the dissociation energy, $D_e^{\rm calc}{=}0.42$\,eV, in E112H
 is close to $D_e^{\rm expt}{=}0.46$\,eV and $D_e^{\rm calc}{=}0.41$\,eV in
 HgH.  These data are quite different from $R_e^{\rm NH}{=}1.829$\,\AA\ and
 $D_e^{\rm NH}{=}0.06$\,eV obtained for E112H within the scalar-relativistic
 Douglas-Kroll approximation [Nakajima and Hirao, Chem.\ Phys.\ Lett., {\bf
 329}, 511 (2000)].  Our results indicate that E112 should not be expected to
 be ``more inert'' than Hg in opposite to the results by other authors.
% and some more calculations with E112 are required to check its chemical
% properties that is important in connection with the planned fundamental
% experiments on the chemical identification of $^{283}$112.
%
\end{abstract}

%\pacs{31.25.Nj, 33.15.-e, 36.90.+f, 31.30.Jv, 31.15.Ar}

\maketitle

%=============================================================================
\paragraph*{Introduction.}
%=============================================================================

%% Paul Scherrer Institut (PSI) Villigen, Switzerland
%% Gesellschaft f{\"u}r Schwerionenforschung Institute (GSI) Darmstadt, Germany
%% Flerov Laboratory of Nuclear Reactions (FLNR) Dubna, Russia
%% Lawrence Berkeley National Laboratory (LBNL) Berkeley, USA
%% Institute of Physical and Chemical Research (RIKEN) Wako-shi, Saitama, Japan

 The superheavy element 112 (eka-Hg) was discovered at GSI (Darmstadt)
 in 1996 within the ``cold'' fusion reaction\cite{Hofmann:96}. 
%??? removed now:
% In the ``hot'' fusion experiment at FLNR (Dubna), a volatile long-lived
% spontaneously fissioning species was observed three years
% later\cite{Oganessian:99b} that was attributed to $^{283}$112 with half-life
%%*** T1/2,alpha > 190 sec.: ***
%%  3\,min 
%%*** T1/2,SF: ***
%  (81$^{+147}_{-32}$) sec.
  The recent observation at FLNR (Dubna) of the ``$\alpha$--SF'' chain,
  attributed to 4\,sec $\alpha$-decay branch of $^{283}$112 followed by a
  0.2\,sec spontaneous fission of $^{279}$110 (Ds)\cite{Oganessian:04}, brought
  up the question, what species was observed 
% in the first FLNR experiment.  
  in the previous ``hot'' fusion FLNR experiment\cite{Oganessian:99b}.  
  Moreover, the production of $^{283}$112 in the reaction of $^{48}$Ca and
  $^{238}$U was not confirmed at LBNL
  (Berkeley)\cite{Loveland:02,Gregorich:05}.  
%??? If there is an exceptionally specific sequence of short $\alpha$-SF
%decays, why chemical experiments are still necessary for indentification?
  However, a very specific decay mode of the short ``$\alpha$--SF'' chain
% or $\alpha-\alpha$ correlation (about 25\% $\alpha$-decay branch for
% $^{279}$Ds was also observed at FLNR) observed at FLNR
  offered a unique chance to unambiguously
% and background free
  identify 
%*** it can be identified only as E112 (ohter will not bee seen)! *** the
%synthesized element
  $^{283}$112 in a chemical experiment.

% Deviant behaviour:
% One element, however, looks set to throw the rulebook away completely. A team
% led by Alexander Yakushev at the Flerov Laboratory of Nuclear Reactions in
% Dubna, Russia, is currently trying to capture element 112. This element is in
% group 12, so it ought to behave like a more volatile version of mercury. But
% relativistic effects are expected to make some of its properties more like
% those of an inert gas such as radon.
%
% Because of strong relativistic contraction of the outermost $7s$ shell have
% unique chemical properties
%
% It was estimated in \cite{Pitzer:75} that the elements 112, 114, and 118
% would be relatively stable as atoms and approach ``inert gas'' (actually
% inert volatile liquid) characteristics (see also \cite{Pitzer:79}).
%
% It was predicted in 1975 by Pitzer \cite{Pitzer:75}
%% Probably,
  To our knowledge, starting from the
%  paper of Pitzer in 1975\cite{Pitzer:75} it was mainly suggested by other
%  authors that E112 behaves rather like a rare gas than Hg, however, Pitzer
%  restricted his consideration by implicit suggestion that E112 forms two
%  $\sigma$ bonds.
   papers of Pitzer\cite{Pitzer:75}
%%<according to Dolg:96:>
   and Fricke\cite{Fricke:75} in 1975 it was mainly suggested by other authors
   that E112 behaves rather like a rare gas than Hg.
% ignoring other possibilities of bonding.  Large difference of Hg and E112 was
% later supported by Kaldor et al. \cite{Eliav:95d,Kaldor:04b}
 In Ref.~\onlinecite{Nash:05}, the
%??? In principle, E112 can be exactly in the middle between Hg and Rn by properties
  confusing
 conclusions about both relative inertness of E112 as compared to Hg and
 similarity of E112 with Hg were made in the abstract and conclusion,
 respectively.
 The first attempt to identify E112 chemically was made at
 FLNR\cite{Yakushev:01,Yakushev:03} but no spontaneous fissions were detected.
 It was interpreted as indication of the Rn-like behavior of E112 as well.

 The chemical experiments on studying properties of E112 are currently under
 way at FLNR\cite{Yakushev:03}, similar work is in progress at GSI and PSI 
 (Villigen)\cite{Eichler:03} that involves the attempt to clarify the recent 
 observation of the decay chains and fission products associated with 
 the production of E114 and E116\cite{Oganessian:04}: being their decay
 product, E112 should be detectable in gas-phase chromatographic experiments.
%??? *** temporarily commented, it should be checked in literature: ***
%  One of the further problems, measuring the atomic masses of the synthesized 
%  isotopes, requires data on the formation probability, in particular, 
%  for E112H and E112H$^+$. 
 The experimental study of superheavy element~(SHE) properties (see
 Refs.~\onlinecite{Schadel:03b,Schadel:06} and references) is very difficult
 because of their short half-lives and extremely small quantities, only single
 atoms are available for research.  In this connection, reliable theoretical
 prediction of their properties based on benchmark {\em ab initio} calculations
 is highly desirable.  As a first step in the extensive study of chemistry of
 E112, bonding in simple diatomic molecules such as E112H should be studied.
 The earlier studies of eka-mercury fluorides\cite{Seth:97} are not so
 sensitive because both mercury and xenon are known to burn in fluorine
 atmosphere. On the other hand, the Hg$_2$, Xe$_2$ and E112$_2$
 dimers\cite{Anton:05,Nash:05} are Van der Waals systems with a small
 dissociation energy.  By contrast, the ground state RnH and XeH molecules are
 not observed in the gas phase, whereas HgH can be obtained by radiofrequency
 discharge in hydrogen and metal vapor (see, e.g.,
 Ref.~\onlinecite{Dufayard:88}).

% The SHE properties can differ from those of the lighter analogues in the
% chemical groups due to very strong relativistic effects on their electronic
% shells. Because of the complicated electronic structure of SHE compounds
% (strong relativistic effects, etc.) and absence of experimental data on their
% properties for calibration, high accuracy and reliability of calculations is
% needed to satisfy the experimental requirements and help in interpreting and
% identifying experimental results. 
 It was shown in Ref.~\onlinecite{Mosyagin:05a} on example of E112 and other
 SHE's that the errors in calculations due to employing the point
%   nuclear model
 nucleus (instead of the realistic Fermi nuclear
%   charge distribution)
 model) reach 0.4~eV for transition energies between low-lying states, whereas
 neglecting the Breit effects leads to the errors up to 0.1~eV.  The
 generalized relativistic effective core potentials (GRECP's)\cite{Titov:99}
 were generated for E112 and other SHE's\cite{Mosyagin:05a} which allow one to
 simulate Breit interaction and Fermi nuclear model by economic way but with
 very high precision\cite{Petrov:04b}.  The accuracy of these GRECP's and of
 the RECP's of other groups was estimated in atomic finite-difference SCF
 calculations with Coulomb two-electron interaction and point nucleus as
 compared to the corresponding all-electron Dirac-Fock-Breit calculations with
 the Fermi nuclear model.  It was justified and checked in
 Refs.~\onlinecite{Petrov:04b,Mosyagin:05a} that the GRECP method allows one to
 carry out reliable calculations of SHE's and their compounds within the level
 of ``chemical accuracy'' (1~kcal/mol, 0.043~eV, or 350~cm$^{-1}$ for valence
 transition energies) when the valence and outer core shells are appropriately
 treated.  Hence, the overall accuracy of calculations in heavy-atom molecules
 is limited, in practice, by current possibilities of the correlation methods
 and codes and not by the GRECP approximation.

 In this paper, we present GRECP calculations of spectroscopic constants for
 the ground states of the E112H molecule and its cation exploiting {\sc molgep}
 code\cite{MOLGEP}. To our knowledge, only three calculations on E112H and
 E112H$^+$ were published\cite{Nakajima:00b,Seth:97,Nash:05}.  In
 Ref.~\onlinecite{Nakajima:00b}, the third order Douglas-Kroll~(DK3) method was
 applied to calculation of E112H and its ions.  The correlations were taken
 into account by the second-order M\o ller-Plesset perturbation theory~(MP2)
 and by the coupled cluster method with single, double (and triple) cluster
 amplitudes, CCSD(T), for 19 external electrons of the E112H molecule.  It is
 not clear from Ref.~\onlinecite{Nakajima:00b} why the correlations with the
 $6p$ shell of E112 are considered but correlations with the $6s$ and $5f$
 shells are not whereas the small value of calculated dissociation energy,
 $D_e^{\rm NH}{=}0.06$\,eV, could be strongly influenced by the latters.  The
 $6s$ and $6p$ shells are closely localized in space, whereas the $5f$ and $6p$
 shells have close orbital energies.  The effect of the finite nuclear size was
 taken into account but the Breit effects and even the spin-orbit~(SO)
 interaction were neglected (i.e., only scalar-relativistic calculations were
 made).  It is clear that both effects are increased with the nuclear charge
 $Z$, therefore, these approximations can be inappropriate for SHE compounds
 even if they are justified for their lighter analogues.
% BSSE for Hirao?

  In Refs.~\onlinecite{Seth:97,Nash:05}, the RECP 
  calculations of E112H$^+$ (but not E112H) were carried out by the
  MP2 and CCSD(T) methods. In Ref.~\onlinecite{Seth:97}, complete active space 
  SCF and multireference configuration interaction (MRCI) calculations of 
  E112H$^+$ were also performed
    \footnote{12 electrons were correlated in MRCI whereas that number
    is not explicitly declared for CCSD(T) calculations with 20-electron PP;
    most likely, all they are correlated.}.
 The SO, finite nuclear size and Breit effects were taken into account at the
 generation stage of the pseudopotential~(PP) of Seth \etal\cite{Seth:97}.  
 It should be noted that the parameters of this PP were
 fitted with the help of the adjustment procedure based on the LS-coupling
%*** at least temporarily enlarged:***
%  scheme (which was, in general,  inappropriate for SHEs).
   scheme (which was, in general, inappropriate for SHE's since the errors of
   these PP's were up to 4~eV for valence energies in our test 
   calculations accounting for the SO interaction non-perturbatively).
 Recently, a new PP for E112 was generated by Seth \etal using the
 $jj$-coupling scheme
    \footnote{P.~Schwerdtfeger, private communication, 2003.},
 but we are not informed about molecular calculations with this PP.
% Moreover, the calculations in \cite{Seth:97} were carried out with the
% spin-averaged PP and then they were corrected for the contribition of the
% spin-orbit interaction.
 The SO interaction was taken into account in the calculations from
 Ref.~\onlinecite{Nash:05}.  However, the author applied the 20-electron RECP
 of Nash \etal\cite{Nash:97}. The Breit effects were not considered at the
 generation stage of this RECP. It is not clear from
 Refs.~\onlinecite{Nash:97,Nash:05} which nuclear model was used there.  In our
 test calculations\cite{Mosyagin:05a}, the errors of this RECP in reproducing
 the results of the all-electron Dirac-Fock-Breit calculations with the Fermi
 nuclear model were up to 1~eV for transition energies between low-lying
 states.  Moreover, the basis set superposition
 errors~(BSSE's)\cite{Gutowski:86,Liu:89} were not estimated in the
 calculations from Ref.~\onlinecite{Nash:05}.

%=============================================================================
\paragraph*{Methods and calculations.}
%=============================================================================

 The GRECP\cite{Titov:99, 
% Titov:00a, 
 Petrov:04b, Mosyagin:05a, Mosyagin:05b}, Fock-space 
%*** CCSD(T) was defined in introduction :::
% relativistic~CC~(RCC) with single and double cluster amplitudes~(SD) 
 relativistic~CC-SD~(RCC-SD)\cite{Eliav:96, Kaldor:04a} 
 and spin-orbit direct configuration
 interaction~(SODCI)\cite{Buenker:99,Alekseyev:04a} methods used for the
 present calculations are well described in literature.  
 The gaussian expansions of our GRECP and $(16,21,16,12,14)/[4,6,4,2,1]$ basis
 set for E112
%(as well as for other elements)
 are available at our website http://www.qchem.pnpi.spb.ru/Basis/.
 In the SODCI calculation, the relativistic scheme of configuration selection
 was applied\cite{Titov:01}.

 Two series of Fock-space RCC-SD calculations were performed for E112H with 
 the GRECP. The ground state of the cation E112H$^+$ served as reference 
 in the first series (denoted by RCC-SD-1), and the Fock-space scheme was
\begin{equation}
   {\rm E112H}^+    \rightarrow  {\rm E112H} ,
 \label{RCC-SD-1}
\end{equation}
 with an electron added in the lowest unoccupied $\sigma$ orbital of E112H$^+$.
 The second series (RCC-SD-2) started from the ground state of the anion
 E112H$^-$ as reference using the Fock-space scheme
\begin{equation}
   {\rm E112H}^- \rightarrow {\rm E112H} \rightarrow {\rm E112H}^+ ,
 \label{RCC-SD-2}
\end{equation}
 with electrons removed from the highest occupied $\sigma$ orbital of E112H$^-$. 
%
%*** Earlier commented to reduce space: ***
 Moreover, the RCC-SD calculations of E112 (to calculate counterpoise
 corrections and $D_e$) were carried out where the $6d^{10} 7s^2$ ground state
 of the E112 atom was used as reference and the Fock-space scheme was
\begin{equation}
   {\rm E112} \rightarrow {\rm E112}^+ ,
 \label{BSSE-RCC-SD}
\end{equation}
 with an electron removed from the $6d$ or $7s$ shell.

 Our test atomic RCC calculations on E112 showed that at least 34 external
 electrons of the atom (occupying the $5f,6s,6p,6d,7s,\ldots$ shells) should be
 correlated and the basis set should include up to $i$-type harmonics ($l{=}6$)
 in order to calculate the excitation and ionization energies with ``chemical
 accuracy''.  Nevertheless, we expect that the contributions of the core
 correlations will be less important for the molecule than for the atom as was
 in our similar calculations\cite{Mosyagin:00,Mosyagin:01b,Mosyagin:05b} on Hg
 and HgH.  This is, in particular, supported
 by a large orbital energy separation between $5p$ and $5d$ shells in Hg
% ($\veps[5p_{3/2}] \approx -5.7$ a.u., $\veps[5d_{3/2}] \approx -1.3$ a.u.)
% ($\veps[5p_{3/2}; 5d_{3/2}] \approx -5.7; -1.3$ a.u.)
  ($\veps[5p_{3/2}; 5d_{3/2}] \approx -2.8; -0.65$ a.u.)
 and by a comparable separation between $6p$ and $6d$ shells in E112
% ($\veps[6p_{3/2}] \approx -4.8$ a.u., $\veps[6d_{3/2}] \approx -1.1$ a.u.)
%  ($\veps[6p_{3/2}; 6d_{3/2}] \approx -4.8; -1.1$ a.u.)
  ($\veps[6p_{3/2}; 6d_{3/2}] \approx -2.4; -0.56$ a.u.)
%
%=====================================================================
%   Hg: s1p1  ee          <r>       |  E112: s1p1  ee           <r>
%---------------------------------------------------------------------
%  5p 1/2   7.245903  0.986595E+00  |  6p 1/2   8.463951  0.100610E+01
%  5p 3/2   5.854771  0.107758E+01  |  6p 3/2   5.015628  0.124085E+01
%  ground   5.7                                 4.8
%  ground   1.3                                 1.1
%  5d 3/2   1.465107  0.142044E+01  |  6d 3/2   1.324044  0.162207E+01
%  5d 5/2   1.312810  0.148309E+01  |  6d 5/2   1.064463  0.176321E+01
%
%  6s 1/2   0.833464  0.273160E+01  |  7s 1/2   1.110533  0.243589E+01
% ====================================================================
%
 Therefore, only 13 external electrons for the E112H molecule (12 electrons for
 E112H$^+$) were correlated in the present calculations.  The calculations with
 the larger number of correlated electrons when SO interaction is explicitly
 treated are rather time-consuming and suggested in future.

%<Hirao carried out calculations only for 19 correlated electrons.>
 In scalar-relativistic CC-SD calculations, we have also estimated (see
 Table~\ref{E112H}) that correlations with the $6p$ shell of E112 give
 relatively small contributions to the spectroscopic constants in E112H and
 E112H$^+$ except for $D_e$ in E112H$^+$.  This cation dissociates to 
% ${\rm E112}^+(6d_{3/2}^4 6d_{5/2}^5 7s_{1/2}^2)~+~{\rm H}(1s^1)$ 
% unlike the HgH$^+$ case.
 ${\rm E112}^+(6d_{3/2}^4 6d_{5/2}^5 7s_{1/2}^2)~+~{\rm H}(1s^1)$\cite{Eliav:95d} 
 in contrast to HgH$^+$, which dissociates to
 ${\rm Hg}^+(5d_{3/2}^4 5d_{5/2}^6 6s_{1/2}^1)~+~{\rm H}(1s^1)$.
 The $6p$ shell is closely localized to the $6d$ shell, therefore, the
 correlations between these shells have to be important for transitions with an
 essential change in the occupation number for the $6d$ shell.  In principle,
 the $D_e$ value for E112H$^+$ can be easily corrected using our atomic RCC
 results for ionization potential of the $6d_{5/2}$ subshell of E112.  However,
 we observed large compensations between contributions accounting for
 correlations with core shells and for basis functions with high angular
 momenta.  Thus, the above ionization potential from calculation with 12
 correlated electrons in the basis including up to $g$-harmonics ($l{=}4$)
 differs from that with 52 electrons and $l$ up to 8 only on
%  +478~cm$^{-1}$ (0.06 eV).
   +773~cm$^{-1}$ (0.10~eV).
%
% As pointed out earlier \cite{Mosyagin,Titov},
% the form~(\ref{GRECP_LS}) of the GRECP operator is optimal
% for calculating states in which changes in occupation numbers
% of outer core shells relative to the state used for the GRECP generation
% are much smaller than 1.
%
% The $D_e$ value corrected                                                                    
% for $5spdf6sp$-core correlations, 
% $hijk$-basis functions, 
% and smoothing the $6d$ spinors is 
% $4.35-(-0.236+0.286+0.040)=4.26$~eV.                                                       
% $5spdf6sp$-core correlations -0.236
% $5spdf6s$-core correlations  +0.072
% $5spdf$-core correlations    +0.054
% $5spd$-core correlations     -0.024
% $hijk$-functions             +0.286
% $ijk$-functions              +0.127
% $jk$-functions               +0.075
% $k$-functions                +0.023

 The calculations were carried out for 15 internuclear distances from 2.3~a.u.\
 to 3.7~a.u.\ with interval of 0.1~a.u. The spectroscopic constants were
 calculated by the Dunham method in the Born-Oppenheimer approximation.
% using the {\sc dunham-spectr} code\cite{Mitin:98}.
 All our RCC and SODCI results reported in Table~\ref{E112H}, were improved
 using counterpoise corrections~(CPC's)\cite{Gutowski:86,Liu:89} calculated
 for the E112 $6d^{10} 7s^2$ state with a ghost H atom. CPC's calculated for the
 ground state of the H atom are about 1~cm$^{-1}$, and are, therefore, ignored.

%=============================================================================
\paragraph*{Results and discussion.}
%=============================================================================

 Our results for the ground states of E112H and E112H$^+$ are collected in
 Table~\ref{E112H}. The corresponding results\cite{Mosyagin:01b,Mosyagin:05b}
 for HgH and HgH$^+$ and the results of other
 groups\cite{Nakajima:00b,Seth:97,Nash:05} for E112H and E112H$^+$ are also
 presented for comparison.  In the \mbox{GRECP/RCC-SD-1} calculations one can
 observe the bond length contraction for E112H and E112H$^+$ on 0.07 and
 0.06~\AA\ with respect to HgH and HgH$^+$.  Detailed comparison of our results
 for HgH and HgH$^+$ with the results of other groups and the experimental data
 can be found in Ref.~\onlinecite{Mosyagin:01b}.

 Our RCC-SD values for spectroscopic constants show considerable differences
 between two Fock-space schemes, GRECP/RCC-SD-1 and GRECP/RCC-SD-2.  Such
 differences are caused by the truncation of the CC operator, they indicate
 significant contributions of the omitted higher-order (triple, etc.) cluster
 amplitudes (HOCA). HOCA influence on the total energies in each point of the
 potential curves were estimated with the help of the configuration interaction
 corrections on HOCA\cite{Isaev:00} calculated as differences in the total
 energies of the SODCI and RCC-SD-1 values. In these calculations,
% 13 electrons were correlated for E112H (12 electrons for E112H$^+$ and E112)
 the same numbers of electrons were correlated as in the above RCC-SD case, but
 a reduced basis set, $[4,4,3,1]$ on E112 and $[3,2]$ on H, was used because
 approaching the full configuration interaction limit in SODCI calculations
 becames too time-consuming for larger basis sets.

 Except for the dissociation limit, HOCA has small effect on E112H$^+$ since
 the cation is a closed shell system.  It is well known that CC approach works
 particularly well for the closed-shell states which is confirmed by the
 comparison of the GRECP/12e-RCC-SD-1 results with and without HOCA correction
 for E112H$^+$. The change in $D_e$ is mainly due to different ionization
 potentials for the $6d_{5/2}$ electron of the E112 atom in the RCC-SD-1 and
 SODCI calculations.  The GRECP/12e-RCC-SD-2 results show considerable
 dictinctions from the results corrected by HOCA because E112H$^+$ is
 calculated in the high Fock-space sector, (2,0), in which some lost of
 accuracy takes place.  The HOCA contribution for E112H is important. Similar
 trend was observed in our GRECP/RCC calculations\cite{Mosyagin:01b} on HgH
 when the effect of the triple cluster amplitudes was taken into account for 13
 electrons that essentially improved the agreement with the experimental data.

 The differences between DK3/CCSD(T)\cite{Nakajima:00b} and our
 GRECP/RCC-SD-1\,+\,HOCA results are small for E112H$^+$, but are essentially
 larger for E112H. In particular, unlike our GRECP/13e-RCC-SD-1\,+\,HOCA bond
 length, the DK3/19e-CCSD(T) value\cite{Nakajima:00b} for E112H is larger than
 the experimental data\cite{Herzberg:50,Stwalley:75,Huber:79,Dufayard:88} for
 HgH. It worth to note that the DK3/CCSD(T) calculations\cite{Nakajima:00b}
 are scalar-relativistic whereas our RCC and SODCI calculations are
% two-component ones
 performed with the spin-dependent GRECP. To check the effect of SO
 interaction, we have also carried out scalar-relativistic CC calculations with
 the spin-averaged GRECP part. The same basis, number of correlated electrons,
 Fock-space schemes, etc.\ were taken as in the RCC calculations.  One can see
 from comparison of our CC-SD and RCC-SD results that the SO effect is small
 for E112H$^+$ (except for $D_e$) but is very essential for E112H. 
% Our GRECP/CC-SD-2 results are in agreement with the DK3/CCSD results
% \cite{Nakajima:00b}.

 The $R_e$ and $w_e$ values for E112H$^+$ by the PP/CCSD(T) method from
 Ref.~\onlinecite{Seth:97} (see the footnote in Table~\ref{E112H})
%??? and Ref.~[29])
 differ from the DK3/18e-CCSD(T) results\cite{Nakajima:00b} by -0.017~\AA\ and
 +45~cm$^{-1}$.  The corresponding PP/CCSD(T)+SO values differ from our
 GRECP/12e-RCC-SD-1~+~HOCA results by -0.023~\AA\ and +126~cm$^{-1}$.  The
 RECP/RCCSD(T) equilibrium distance for E112H$^+$ from
 Ref.~\onlinecite{Nash:05} differs from our GRECP/RCC-SD-1\,+\,HOCA result by
 +0.04~\AA\ and even more from the results of other
 groups\cite{Seth:97,Nakajima:00b}.  The difference between the RECP/RCCSD(T)
 equilibrium distance calculated in Ref.~\onlinecite{Nash:05} for HgH$^+$ and
 the experimental datum is -0.04~\AA.  Thus, one can observe the larger bond
 length for E112H$^+$ by 0.03~\AA\ in comparison with HgH$^+$ in the
 RECP/RCCSD(T) calculations\cite{Nash:05} (in contrast to our results).

% Question/Suggestion only: May be it is worth to start the RCC calculations
% on E112H with the 6p (6s) shell correlated?

 One, however, can expect rather some increase in $R_e$ for E112H$^+$ and
 decrease for E112H when accounting for correlations with the innermore shells.
 In our scalar-relativistic calculations, some small increase in $D_e$ was
 observed with enlarging the basis set.

%=============================================================================
\paragraph*{Conclusions.}
%=============================================================================

 It is well known that properties of SHE's are somewhat different from those of
 their lighter analogues due to very strong relativistic effects first of 
 all (see Ref.~\onlinecite{Kaldor:04a} and references).
 Therefore, even those approximations which work well for the lighter 
 analogues (neglecting the SO interaction for $\Sigma$ states, the effective 
 state of a considered atom in a molecule, the preferred valency, etc.) should 
 not be used for SHE's without serious checking and analyzing. 
 The calculated equilibrium distance, $R_e$,
 in E112H is notably smaller than that in HgH. Therefore, one can also expect
% smaller effective raduis for the E112 atom in its other compounds in
  smaller bond lengths for the other E112 compounds in
 comparison with the Hg ones.  There is a long-term discussion in scientific
 community whether E112 will behave like Hg or Rn. 
% In DiRef database there are articles on observation of
%    XeH compounds in matrices of rare gases.
 The ground state RnH and XeH molecules are not observed in the gas phase.  Our
 calculations for the E112H molecule do not predict large dissociation energy,
 $D_e$, but it is yet close to that of HgH.  
%<May be better to remove it. In opposite case, we should also cite>
%<here Anton, Nash calculations on E112_2>
%  Our preliminary 
%  calculations\footnote{A.N.Petrov, et al., in preparation.} 
%  also predict the similar properties for Hg$_2$ and E112$_2$.
 Therefore, we believe that the
%??? The last phrase is not clear
%***??? to be inserted later:***
%<May be better to insert now because of the first referee remark.>
  singly-valent ground state E112 compounds will rather resemble
% the Hg ones than will have the noble gas behavior.
% the Hg-like than the noble gas behavior.
% the Hg compounds than E112 will have the noble gas behavior.
  the Hg compounds than the noble gas patterns that is also supported by
  calculations of other E112
  compounds\cite{Pershina:02a,Zaitsevskii:06a,Petrov:06a}.

%temporary deleted ***??? to be inserted later:***
%  However, calculations for a wide class of its compounds, in particular, of
%  dimer E112$_2$ are important to make a definitive conclusion about the
%  valence properties of E112.
%
% It is known that Hg is a volatile liquid and Rn is a gas.  One can expect
% that E112 will be also volatile.

%??? to add:
% and some more calculations with E112 are required to check its volatility
% that is important in connection with the planned fundamental experiments on
% the chemical identification of $^{283}$112.

%=============================================================================
\paragraph*{ACKNOWLEDGMENTS.}
 
%*** maybe better to move that to references: ***
% We are grateful to R.J.Buenker and his colleagues H.-P.Liebermann and
% A.B.Alekseyev for giving us the new version of {\sc sodci} code and to
% U.Kaldor and E.Eliav for the {\sc rcc-sd} code which were used in the present
% calculations.
 The {\sc sodci} code of R.J.Buenker and his colleagues H.-P.Liebermann and
 A.B.Alekseyev together with the {\sc rcc-sd} code of U.Kaldor, E.Eliav, and
 A.Landau were used in our calculations.
 We are grateful to A.Zaitsevskii and Yu.Tchuvil'sky for discussions and
 corrections in the paper.
 The present work is supported by the RFBR grant 03--03--32335.  N.M.\ thanks
 Russian Science Support Foundation. 
% A part of the CI calculations was performed on computers of Boston University
% in the framework of the MARINER project.

%=============================================================================

\clearpage

\bibliographystyle{apsrev}

\bibliography{bib/JournAbbr,bib/Titov,bib/TitovLib,bib/Kaldor,bib/Isaev,bib/TitovAbs}

%=============================================================================

\squeezetable
\renewcommand{\baselinestretch}{1}
\begin{table*}%[t]
\protect\renewcommand{\baselinestretch}{1.5}
\caption{Spectroscopic constants of the ground states of 
   the E112H molecule and the E112H$^+$ ion from 
%  13- and 12-electron 
   two-component RCC-SD and scalar-relativistic CC-SD calculations with GRECP 
   in the H $(8,4,3)/[4,2,1]$ ANO and E112 $(16,21,16,12,14)/[4,6,4,2,1]$
   basis set.  
   Our corresponding results for HgH and HgH$^+$
   and the results of other groups for E112H and E112H$^+$
   are also presented for comparison.
%***To save space:***
%   All our results are corrected by CPCs
%   calculated for the ground states of the E112 and Hg atoms.
%
   $R_e$ is in \AA, $D_e$ in eV, $Y_{02}$ in $10^{-6}$~cm$^{-1}$, 
   other values in cm$^{-1}$.
}
%
%???
%\bigskip
% \medskip
%\smallskip
\begin{tabular}{llccccccc}
\hline
\hline
Molecula   & Method                                & $R_e$ & $w_e$ & $D_e$ & $B_e$ & $w_e x_e$ & $\alpha_e$ & $-Y_{02}$ \\
\hline                                                                                                             
\multicolumn{9}{c}{Our calculations:}\\
  HgH$^+$  & GRECP/12e-RCC-SD-1                    & 1.596 & 2037  & 2.67  & 6.60  &  ~39      &     0.200  &      279  \\
  HgH$^+$  & GRECP/12e-RCC-SD(T)-1                 & 1.599 & 2013  & 2.68  & 6.58  &  ~41      &     0.208  &      282  \\
\cline{1-2}
  HgH$^+$  & Experiment\cite{Herzberg:50,Huber:79} & 1.594$\pm$0.000 & 2031$\pm$3 & (2.75$\pm$0.36)$^{\rm a}$ & 6.61$\pm$0.00 & 44$\pm$3 & 0.206$\pm$0.000 & 285$\pm$0 \\
%\cline{1-2}%--------------------------------------------------------------------------------------------------------------
 \hline
\multicolumn{9}{c}{Our calculations:}\\
 E112H$^+$ & GRECP/20e-CC-SD-1                     & 1.537 & 2587  & 4.60  & 7.10  &  ~46      &     0.198  &      215  \\
 E112H$^+$ & GRECP/20e-CC-SD-2                     & 1.531 & 2681  & 4.46  & 7.15  &  ~35      &     0.168  &      205  \\
 E112H$^+$ & GRECP/18e-CC-SD-1                     & 1.537 & 2588  & 4.61  & 7.10  &  ~47      &     0.198  &      215  \\
 E112H$^+$ & GRECP/18e-CC-SD-2                     & 1.531 & 2680  & 4.46  & 7.16  &  ~35      &     0.169  &      205  \\
 E112H$^+$ & GRECP/12e-CC-SD-1                     & 1.535 & 2590  & 4.96  & 7.12  &  ~47      &     0.200  &      216  \\
 E112H$^+$ & GRECP/12e-CC-SD-2                     & 1.527 & 2679  & 4.75  & 7.19  &  ~37      &     0.175  &      208  \\
 E112H$^+$ & GRECP/12e-RCC-SD-1                    & 1.537 & 2569  & 3.96  & 7.11  &  ~47      &     0.201  &      218  \\
 E112H$^+$ & GRECP/12e-RCC-SD-2                    & 1.519 & 2752  & 3.80  & 7.28  &  ~45      &     0.187  &      204  \\
%E112H$^+$ & GRECP/12e-RCC-SD-1~+~HOCA (Old SODCI) & 1.538 & 2546  & 4.36  & 7.09  &  ~47      &     0.203  &      221  \\
%E112H$^+$ & GRECP/12e-RCC-SD-1~+~HOCA (11 points) & 1.540 & 2544  & 4.35  & 7.08  &  ~45      &     0.196  &      220  \\
%E112H$^+$ & GRECP/12e-RCC-SD-1~+~HOCA (13 points) & 1.540 & 2545  & 4.35  & 7.07  &  ~43      &     0.187  &      219  \\
%E112H$^+$ & GRECP/12e-RCC-SD-1~+~HOCA             & 1.540 & 2547  & 4.35  & 7.08  &  ~45      &     0.195  &      220  \\
 {\bf E112H$^+$} & {\bf GRECP/12e-RCC-SD-1~+~HOCA}             & {\bf 1.540} & {\bf 2547}  & {\bf 4.35}  & {\bf 7.08}  &  {\bf ~45}      &     {\bf 0.195}  &      {\bf 220  }\\
\cline{1-2}%--------------------------------------------------------------------------------------------------------------
\multicolumn{9}{c}{Other groups' calculations:}\\
%A: 275-->296 : B_e and w_e x_e are decreased in 1.00026 times,                                                    
%               w_e is decreased in 1.00013 times                                                                  
%Hirao A=275   (corrected for A=296 B_e=7.184 and 7.149)                                                                   
%E112H$^+$ & DK3/18e-CCSD \cite{Nakajima:00b}      & 1.528 & 2621  &       & 7.186 &           &            &           \\
%E112H$^+$ & DK3/18e-CCSD(T) \cite{Nakajima:00b}   & 1.532 & 2595  &       & 7.151 &           &            &           \\
 E112H$^+$ & DK3/18e-CCSD\cite{Nakajima:00b}       & 1.528 & 2621  &       & 7.18  &           &            &           \\
 E112H$^+$ & DK3/18e-CCSD(T)\cite{Nakajima:00b}    & 1.532 & 2595  &       & 7.15  &           &            &           \\
%Dolg A=275                                                                                                             
%E112H$^+$ & PP/CCSD(T) \cite{Seth:97}$^{\rm b}$   & 1.515 & 2640  & 5.154 &       &  ~51      &            &           \\
%E112H$^+$ & PP/MRCI+SO \cite{Seth:97}$^{\rm b}$   & 1.503 & 2620  & 3.856 &       &           &            &           \\
%E112H$^+$ & PP/CCSD(T)+SO \cite{Seth:97}$^{\rm b}$& 1.517 & 2673  & 4.094 &       &  ~52      &            &           \\
 E112H$^+$ & PP/CCSD(T)\cite{Seth:97}$^{\rm b}$    & 1.515 & 2640  & 5.15  &       &  ~51      &            &           \\
 E112H$^+$ & PP/MRCI+SO\cite{Seth:97}$^{\rm b}$    & 1.503 & 2620  & 3.86  &       &           &            &           \\
 E112H$^+$ & PP/CCSD(T)+SO\cite{Seth:97}$^{\rm b}$ & 1.517 & 2673  & 4.09  &       &  ~52      &            &           \\
%Nash A=0 ?
 E112H$^+$ & RECP/RCCSD(T)\cite{Nash:05}$^{\rm b}$ & 1.583 &       & 3.50  &       &           &            &           \\
%E112H$^+$ & RECP/RCCSD\cite{Nash:05}$^{\rm b}$    & 1.543 &       & 3.41  &       &           &            &           \\
%                                                                                                                               
\hline%===================================================================================================================
\hline
\multicolumn{9}{c}{Our calculations:}\\
  HgH      & GRECP/13e-RCC-SD-1                    & 1.709 & 1575  &  0.35 & 5.76  &  ~56      &     0.262  &      312  \\
  HgH      & GRECP/13e-RCC-SD(T)-1                 & 1.738 & 1395  &  0.41 & 5.56  &  ~83      &     0.348  &      363  \\
\cline{1-2}
  HgH      & Experiment\cite{Herzberg:50,Stwalley:75,Dufayard:88}  & 1.738$\pm$0.003 & 1403$\pm$18 & 0.46$\pm$0.00 & 5.57$\pm$0.02  & 98$\pm$23 & 0.337$\pm$0.067 & 345$\pm$1 \\
  HgH      & Experiment\cite{Huber:79}             & [1.766]$^{\rm c}$ & [1203]$^{\rm c}$ & ~0.46 & [5.39]$^{\rm c}$ & & & [395]$^{\rm c}$ \\
%\cline{1-2}%--------------------------------------------------------------------------------------------------------------
 \hline
\multicolumn{9}{c}{Our calculations:}\\
 E112H     & GRECP/21e-CC-SD-1                     & 1.742 & 1438  & -0.03 & 5.53  &  113      &     0.409  &      340  \\
 E112H     & GRECP/21e-CC-SD-2                     & 1.801 & 1104  & -0.02 & 5.15  &           &            &           \\
% w_e x_e, \alpha_e, -Y_02 are not stable for decrease in the number of points
%E112H     & GRECP/21e-CC-SD-2                     &       &       &       &       &  215      &     0.702  &      530  \\
 E112H     & GRECP/19e-CC-SD-1                     & 1.741 & 1439  & -0.03 & 5.53  &  113      &     0.409  &      340  \\
 E112H     & GRECP/19e-CC-SD-2                     & 1.801 & 1102  & -0.02 & 5.15  &           &            &           \\
 E112H     & GRECP/13e-CC-SD-1                     & 1.746 & 1402  & -0.03 & 5.50  &  119      &     0.429  &      354  \\
 E112H     & GRECP/13e-CC-SD-2                     & 1.808 & 1038  & -0.05 & 5.10  &           &            &           \\
% w_e x_e, \alpha_e, -Y_02 are not stable for decrease in the number of points
%E112H     & GRECP/13e-CC-SD-2                     &       &       &       &       &  274      &     0.860  &      635  \\
 E112H     & GRECP/13e-RCC-SD-1                    & 1.638 & 1859  & 0.36  & 6.25  &  ~95      &     0.338  &      288  \\
 E112H     & GRECP/13e-RCC-SD-2                    & 1.663 & 1649  & 0.32  & 6.06  &  123      &     0.425  &      340  \\
%
%E112H     & GRECP/13e-RCC-SD-1~+~HOCA (Old SODCI) & 1.668 & 1738  &  0.41 & 6.02  &  168      &     0.417  &      304  \\
%E112H     & GRECP/13e-RCC-SD-1~+~HOCA (11 points) & 1.662 & 1774  &  0.42 & 6.07  &  134      &     0.391  &      294  \\
%E112H     & GRECP/13e-RCC-SD-1~+~HOCA (13 points) & 1.660 & 1840  &  0.42 & 6.07  &  181      &     0.392  &      279  \\
%E112H     & GRECP/13e-RCC-SD-1~+~HOCA             & 1.662 & 1800  &  0.42 & 6.07  &  152      &     0.385  &      287  \\
{\bf E112H}     & {\bf GRECP/13e-RCC-SD-1~+~HOCA}             & {\bf 1.662} & {\bf 1800}  &  {\bf 0.42} & {\bf 6.07}  &  {\bf 152}      &     {\bf 0.385}  &      {\bf 287}  \\
\cline{1-2}%--------------------------------------------------------------------------------------------------------------
\multicolumn{9}{c}{Other groups' calculations:}\\
%Hirao A=275   (corrected for A=296  B_e= 5.051 and 5.016)                                                           
%E112H     & DK3/19e-CCSD \cite{Nakajima:00b}      & 1.823 & ~991  & 0.036 & 5.052 &           &            &           \\
%E112H     & DK3/19e-CCSD(T) \cite{Nakajima:00b}   & 1.829 & 1007  & 0.057 & 5.017 &           &            &           \\
 E112H     & DK3/19e-CCSD\cite{Nakajima:00b}       & 1.823 & ~991  &  0.04 & 5.05  &           &            &           \\
 E112H     & DK3/19e-CCSD(T)\cite{Nakajima:00b}    & 1.829 & 1007  &  0.06 & 5.02  &           &            &           \\
\hline
\hline
\end{tabular}
\label{E112H}
\begin{flushleft}
\noindent $^{\rm a}$ Cited in Refs.~\onlinecite{Herzberg:50,Huber:79}
                    as uncertain.

\noindent $^{\rm b}$ Note that 
  the RCCSD(T), RMRCI+SO, RCCSD(T)+SO values from Ref.~\onlinecite{Seth:97} and
  the CCSD(T) values from Ref.~\onlinecite{Nash:05}
 are listed. The acronyms for these
  calculations (the last one is in the $jj$-coupling scheme, the other ones are
  scalar-relativistic where the second and third ones are corrected for the SO
  effects) 
 are redefined in accordance with the other notations of the present paper.

\noindent $^{\rm c}$ Cited in Ref.~\onlinecite{Huber:79} 
                     as corresponding to the zero vibrational level.
\end{flushleft}

\end{table*}

\end{document}